\documentclass[10pt, twocolumn, fleqn]{article}
\usepackage[utf8]{inputenc}
\usepackage{graphicx}
\usepackage[fleqn]{amsmath}
\usepackage[version=4]{mhchem}
\usepackage{siunitx}
\usepackage{longtable,tabularx}
\usepackage{nccmath}
\usepackage[superscript]{cite}
\usepackage[margin=1in]{geometry}
\usepackage[toc,page]{appendix}
\usepackage{setspace}
\usepackage{amssymb}
\usepackage{listings}
\usepackage{afterpage}
\usepackage{titlesec}
\usepackage[labelfont=bf]{caption}
\usepackage{fancyhdr}
\usepackage{pgf}
\usepackage{float}
\usepackage{microtype}
\usepackage{subcaption}
\usepackage[parfill]{parskip}
\usepackage{authblk}
\usepackage{rotating}

\usepackage{amsfonts}
 \newtheorem{assumption}{Assumption}
  \newtheorem{lemma}{Lemma}
 
 \newtheorem{remark}{Remark}
 \newtheorem{problem}{Problem}

 \usepackage{algorithm}
\usepackage[noend]{algpseudocode}
\usepackage{graphicx}

\usepackage{enumitem}
\usepackage{multicol}

\title{Control of Small Spacecraft by Optimal Output Regulation: A Reinforcement Learning Approach}
\author[1]{Joao Leonardo Silva Cotta}
\author[1]{Omar Qasem}
\author[1]{Paula do Vale Pereira}
\author[1]{Hector Gutierrez}
\affil[1]{Florida Tech}
\date{June 2023}

\titleformat{\section}
{\normalfont\normalsize\bfseries}{\thesection}{0.5em}{}
\titleformat{\subsection}
{\normalfont\normalsize\bfseries\em}{\thesubsection}{0.5em}{}
\titleformat{\subsubsection}
{\normalfont\normalsize\bfseries\em}{\thesubsubsection}{0.5em}{}
\usepackage[toc,acronym]{glossaries} 
\newglossary[nlg]{notation}{not}{ntn}{Notation} 
\makeglossaries
\titleformat{\section} [block]{\normalfont\bfseries}{}{0em}{\MakeUppercase}

\begin{document}

\twocolumn[
  \begin{@twocolumnfalse}
    \begin{flushright}
    \Large
    \textbf{[SSC23-WP1-21]} 
    \end{flushright} 
    \vspace{1cm}
    \begin{centering}      
    \large
    \textbf{Control of Small Spacecraft by Optimal Output Regulation: A Reinforcement Learning Approach}\\
    \vspace{1.5em}
    \normalsize
    Joao Leonardo Silva Cotta, Omar Qasem, Paula do Vale Pereira, Hector Gutierrez\\
    Florida Institute of Technology, Department of Aerospace, Physics, and Space Sciences\\
    150 W. University Blvd, Melbourne, FL 32901 \\
    oqasem2021@my.fit.edu \\
    \vspace{1.5em}

    \textbf{ABSTRACT} \\ 
    \vspace{0.3cm}
    \end{centering}
            \normalsize
            The growing number of noncooperative flying objects has prompted interest in sample-return and space debris removal missions. Current solutions are both costly and largely dependent on specific object identification and capture methods. In this paper, a low-cost modular approach for control of a swarm flight of small satellites in rendezvous and capture missions is proposed by solving the optimal output regulation problem. By integrating the theories of tracking control, adaptive optimal control, and output regulation, the optimal control policy is designed as a feedback-feedforward controller to guarantee the asymptotic tracking of a class of reference input generated by the leader. The estimated state vector of the space object of interest and communication within satellites is assumed to be available. The controller rejects the nonvanishing disturbances injected into the follower satellite while maintaining the closed-loop stability of the overall leader-follower system. The simulation results under the Basilisk-ROS2 framework environment for high-fidelity space applications with accurate spacecraft dynamics, are compared with those from a classical linear quadratic regulator controller, and the results reveal the efficiency and practicality of the proposed method. \\
  \end{@twocolumnfalse}
]

\section*{Introduction}

In the growing space exploration context, noncooperative objects, such as space debris, have become a leading interest regarding collision avoidance and the challenges associated with identifying, tracking, and removing or retrieving such objects. Available solutions are often heavily dependent on high-cost complex systems that lack modularity, being designed for a specific object size range, thus limiting their applicability and efficacy. This investigation aims to address the presented issues through a low-cost approach that leverages the usage of a multitude of small satellites with a cooperative optimal output regulation control to rendezvous and capture the space debris of interest.

Ballistic-controlled differential drag controllers have been explored for maneuvering in low earth orbit (LEO) as a low-cost alternative, particularly for formation flying and orbit maintenance. Thus, many successful applications can be found in the literature for small satellites and CubeSats since the limitations surrounding mounting thrusters or fuel expenditure make attitude controllers an attractive alternative \cite{harris1,PEREZ2013196}. The method essentially consists of actuating the spacecraft's surface panels to control the multiple ballistic coefficients. However, modulating the space vessel with multiple control surfaces may represent additional costs and increase the mission complexity as controlling the spacecraft in between moving these surfaces can be significantly challenging, especially when dealing with discrete-attitude-mode ``bang-bang'' orbit control \cite{horse1}. In this perspective, recent innovative differential-drag control techniques have been presented for CubeSats, such as continuous differential-drag control using single-axis attitude-driven orbit control that has gained popularity as the online optimizer and compensator for the QB50 constellation\cite{DELLELCE2015112}. In addition, recent developments adapting coupled rotational-transitional control using dual quaternion representation for space rendezvous missions have gained notoriety as the usage of electric and solar sails increases in small satellites\cite{harris1,felipeta1}. Furthermore, the work presented by \textit{Harris et al.} has explored the linear coupled attitude-orbit control through aerodynamic drag; however, the development is centered on approaching the problem using a linear Quadratic Regulator (LQR), requiring an accurate dynamic model of the spacecraft system being analyzed. To avoid the prior knowledge of the system's dynamics, adaptive control techniques in the context of differential-drag control have been previously explored, but, to the best of the authors' knowledge, none have overcome the need for a dynamic model while maintaining optimal performance, rectifying the importance of the development presented in this investigation \cite{PEREZ2013196}.

\subsection*{LQR Limitations in Space Applications}

The limitations surrounding the well-known LQR control have been previously explored in the literature and are mainly due to the necessity of prior knowledge of the system's dynamics to solve the optimal control problem. When it comes to space systems and missions, most of this information can be hindered given the mission's nature or is costly to obtain. For instance, to deploy a swarm of CubeSats using LQR, performing an accurate system identification is significantly complex and not a trivial task, not only due to the non-linearities inherent in the space environment but also the unknown dynamic characteristics that come from gravity, drag, magnetic field, space debris, and radiation.

In addition, the LQR has known difficulties when dealing with multi-agent systems, given its linear control structure and the challenges of managing the interdependence among the swarm members. Therefore, these cooperative systems require more robust control. A potential alternative for the output regulation theory concerning the nonvanishing disturbances problem would be using adaptive control combined with reinforcement learning techniques to handle some of the previously introduced issues.

Generally speaking, the output regulation problem has gained the consideration and attraction of a wide audience in control systems society since it is a general mathematical formulation to tremendous control problems applications in engineering, biology, satellite clustering and other disciplines; see, for instance, \cite{bonivento2001output, huang2004nonlinear, isidori2003robust, sontag2003adaptation, trentelman2002control, Gao2016Auto,QasemCDC2022,QasemTIE2023} and many references therein. The linear output regulation problem is mainly concerned in designing a control policy to achieve asymptotic tracking of a class of reference inputs, in addition to rejecting nonvanishing disturbances, in which both the reference signals and the disturbances are generated by a class of autonomous systems, named exosystems. The output regulation problem is essentially solved using either the feedback-feedforward method or the internal model principle. In this work, we focus on solving the output regulation problem in the feedback-feedforward scheme.

In this paper, the tracking control problem is investigated by solving the optimal output regulation problem and developing an optimal differential drag feedback-feedforward control policy for a follower spacecraft on based on adaptive dynamic programming (ADP) techniques rooted in reinforcement learning to establish a flexible control law that does not require detailed information on the dynamic model of the hive. ADP techniques have been widely investigated in the literature \cite{Frank2012optimal,lewis2009reinforcement,AbuKhalaf2005,Abukhalaf2006,QasemTASE2023,QasemTIE2023,qasem2023autonomous,qasem2023improving,Qasem2021HI,QasemCDC2022}. ADP methods are mainly built upon policy iteration (PI) and value iteration (VI). In this work, VI is to be considered since a stabilizing control policy is not required to initiate the learning process. In practice, the proposed method can be introduced into most modern spacecraft systems that employ a message-based command and data handling framework, utilizing data packets from each subsystem to inform the controller.

The newly introduced optimal control policy combines tracking control, adaptive control, and output regulation. The methodology is centered on a differential drag feedback-feedforward controller that guarantees the asymptotic tracking of a class of reference input generated by the lead satellite. The controller rejects the nonvanishing disturbances injected into the followers while maintaining the closed-loop stability of the overall multi-agent system. The approach is validated and compared to a classical Linear Quadratic Regulator using the high-fidelity astrodynamics framework Basilsik combined with ROS2. Furthermore, another prominent aspect of this paper is to develop and implement a controller that can be used in a real mission scenario and easily implemented on most current small satellite systems using a messaging framework. These efforts shall increase each subsystem's modularity and architecture independence, providing the possibility to make changes to the controller objective and propose based on the mission statement.

The contributions of this paper are summarized by the following:
(i) We consider the autonomous formation flying of a swarm of small satellites  in  a rendezvous and capture mission with non-cooperative debris under the effects of atmospheric drag.
(ii) The Clohessy-Wiltshire equations are considered, wherein the optimal feedback-feedforward control gain matrices are obtained using value iteration.
(iii) It is shown that the capture of the debris by the swarm lead is perfectly achieved in an optimal sense by considering reinforcement learning with the output regulation problem.
(iv) To the best of our knowledge, this work is the first to consider ADP strategies and the output regulation concept rooted in reinforcement learning in high-fidelity autonomous rendezvous and capture mission applications to regulate the relative position of the center of a small satellite formation flight.

\subsubsection*{\textbf{Notations.}}

\ The operator $|\cdot|$ represents the Euclidean norm for vectors and the induced norm for matrices. $\mathbb{Z}_{+}$ denotes the set of nonnegative integers. The Kronecker product is represented by $\otimes$, and the block diagonal matrix operator is denoted by $\textrm{bdiag}$. 
$I_n$ denotes the identity matrix of dimension $n$ and $0_{n\times m}$ denotes a $n\times m$ zero matrix. 
$\text{vec}(A) = [a_1^\textrm{T},a_2^\textrm{T},...,a_m^\textrm{T}]^\textrm{T}$, where $a_i \in \mathbb{R}^n$ is the $i^{\text{th}}$ column of $A \in \mathbb{R}^{n\times m}$. 
For a matrix $P=P^\textrm{T} \in \mathbb{R}^{m\times m},$ $\text{vecs}(P)=\left[p_{11},2p_{12},\ldots,2p_{1m},p_{22}, \right. \\ \left. 2p_{23},\ldots,2p_{m-1,m},p_{mm}\right]^\textrm{T} \in \mathbb{R}^{\frac{1}{2}m(m+1)}.$ 
$P\succ(\succeq)0$ and $P\prec(\preceq)0$ denote the matrix $P$ is positive definite (semidefinite) and negative definite (semidefinite), respectively. For a column vector $v\in \mathbb{R}^n$, ${\rm vecv}(v)$$=[v_1^2,v_1 v_2,\ldots,v_1v_n,v_2^2, \\ v_2v_3,\ldots,v_{n-1}v_n,v_n^2]^\textrm{T} \in \mathbb{R}^{\frac{1}{2}n(n+1)}$. For a matrix $A\in \mathbb{R}^{n\times n}$, $\sigma(A)$ denotes the spectrum of $A$. For any $\lambda \in \sigma(A)$, $\text{Re}(\lambda)$ represents the real part of the eigenvalue $\lambda$.

\section*{Objective and Methodology}

This paper aims to study and analyze the rendezvous and capture scenario of a swarm of small satellites and a noncooperative spacecraft. The process is done with guaranteed stability and maintaining the asymptotic tracking of the swarm lead satellite to the debris. Moreover, the method should not rely on prior knowledge of the physics of the system. Therefore, the learning method is carried out by the adaptive optimal output regulation, in which the optimal control policy is learned using ADP.

To begin with, consider the following continuous-time linear system described by 
\begin{align}\label{eq: exosystem}
\dot{v}&=Ev,\\\label{eq: x-system}
    \dot{x}&=Ax+Bu+Dv\\\label{eq: e system}
    e&=Cx+Fv,
\end{align}
where the vector $x\in\mathbb{R}^{n}$ is the state, $u\in\mathbb{R}^{m}$ is the control input, and $v\in\mathbb{R}^q$ stands for the exostate of an autonomous system \eqref{eq: exosystem}. The vector $e\in\mathbb{R}^{p}$ represents the output tracking error. The matrices $A\in\mathbb{R}^{n\times n}$, $B\in\mathbb{R}^{n\times m}$, $D\in\mathbb{R}^{n\times q}$, $C\in\mathbb{R}^{p\times n}$, and $F\in\mathbb{R}^{p\times q}$ are real matrices with the pair $(A,B)$ are assumed to be unknown.

In our case, the relative motion between a 'Deputy' -- the swarm lead satellite --  and a 'Chief' -- debris--  which are in close vicinity of each other is defined by the Clohessy-Wiltshire (CW) including atmospheric drag equations \cite{schaub2003analytical,Silva2008}. Relative accelerations in Cartesian coordinates are given by:
\begin{align}
\begin{split}\label{eq: CW 1}
0&=\Ddot{\textbf{x}} - 2\dot {\textbf{y}} \bar{n} - 3\bar{n}^2\textbf{x} + \frac{\beta_d P_d \bar {n} r_c}{2} \dot{\textbf{x}}
\end{split}
\\
\begin{split}\label{eq: CW 2}
0&=  \ddot{\textbf{y}}+2\bar{n}\dot{\textbf{x}}+\beta_d P_d \bar {n} r_c \dot{\textbf{y}} \\
&+ \frac{\bar{n}^2{r_c}^2(\beta_c P_c - \beta_d P_d)}{2}
\end{split}
\\
\begin{split}\label{eq: CW 3}
0&=\ddot{\textbf{z}}+\bar{n}^2\textbf{z} + \frac{\beta_d P_d \bar {n} r_c}{2} \dot{\textbf{z}}
\end{split}
\end{align}
where $(\textbf{x},\textbf{y},\textbf{z})$ represent the relative position of the two satellites in the orthogonal Cartesian coordinate system, $r_c$ is the equivalent radius of the chief, and $\bar n$ is the mean orbital rate. The vector components are taken in the rotating chief Hill frame. The advantage of using Hill frame coordinates is that the physical relative orbit dimensions are immediately apparent from these coordinates. The $(\textbf{x},\textbf{y})$ coordinates define the relative orbit motion in the chief orbit plane. The $\textbf{z}$ coordinate defines any motion out of the chief orbit plane. In addition, $\beta$ is the ballistic coefficient due to a collection of \textit{\textbf{n}} flow exposed panels \cite{sutton2008}, and $P_d$ and $P_c$ represent the altitude-dependent atmospheric density of the deputy and chief, respectfully.

Note that for an exponential atmosphere, the relation is given by \cite{harris1}:
\begin{align}\label{eq:atmsdensity}
    P_d = P_c\ e^{-x/H} \approx P_c (1 - \textbf{x}/H)
\end{align}
which holds within one atmospheric scale height of the center of the swarm formation position.
The following assumptions are taken into account such that \eqref{eq: CW 1}-\eqref{eq: CW 3} can be held.
\begin{assumption} The relative distance between the chief and the deputy is much smaller than the orbit radius $r$.
\end{assumption}
\begin{assumption}
The relative orbit is assumed to be circular.
\end{assumption}

  \begin{figure*}
    \centering
    \includegraphics[width = 0.71\textwidth
    ]{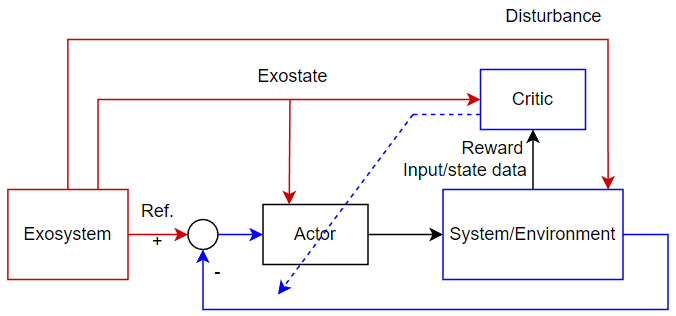}
    \captionsetup{font=small}
    \caption{The framework of the learning-based adaptive optimal output regulation}
    \label{fig: actor critic diagram}
\end{figure*}
To begin with, we define the state space vector $x$ as $x=\left[\textbf{x},~\textbf{y},~ \textbf{z},~\dot{\textbf{x}},~\dot{\textbf{y}},~ \dot{\textbf{z}}\right]^\textrm{T}$
We can then write eqs \eqref{eq: CW 1}-\eqref{eq: CW 3} in the form of \eqref{eq: exosystem}-\eqref{eq: e system} and further define the system in terms of variation from an arbitrary reference attitude $\mathbf{\sigma_p}$ as the control input of the system, having matrix $B$ being defined as:
\begin{align}
    B&=\begin{bmatrix}
  {0}_{3\times3}\\
     {0}_{1\times3}\\
        \frac{1}{2}\bar n^2 {r_c}^2 P_d \ \partial \beta_d/\partial {\mathbf{\sigma_p}}\\
       {0}_{1\times3}
\end{bmatrix},\\
    u&=\begin{bmatrix}
        \mathbf{\sigma_{P,1}}\\
        \mathbf{\sigma_{P,2}}\\
        \mathbf{\sigma_{P,3}}
\end{bmatrix}
\end{align}
where $\partial \beta_d/\partial {\sigma_p}$ is the sensitivity of the ballistic coefficient on attitude. As shown by \textit{Harris et al.}, the sensitivity can be defined as follows:
\begin{align}
    \frac{\partial \beta_d}{\partial \sigma_p} = \frac{1}{\bar m} \sum_{i=1}^{n} 4 C_{D;i} A_i  \mathbf{\hat{n_i}}^T \mathbf{\hat{q}}\times
\end{align}
here, $\bar{m}$ represents the mass, $C_D$ is the drag coefficient, $A_i$ is the area per-facet, $\mathbf{\hat{q}}\times$ is the intermediate vector described by $\mathbf{BN}(\mathbf{\sigma_p}\mathbf{\hat v})$, and finally $\mathbf{N}$ is the planet-centered inertial coordinate frame:
\begin{align}
    \mathbf{N} = <0, \mathbf{\hat n_1}, \mathbf{\hat n_2}, \mathbf{\hat n_3}>
\end{align}
Therefore, the system dynamics equations in terms of the sensitivity can be expressed for an equilibrium point as \cite{harris1,harris2}:
\begin{align}
    \frac {\partial {\Ddot{\textbf{x}}}}{\partial \mathbf{\sigma_p}} &= \frac{-1}{2} P_d r_c \bar n \dot{\textbf{x}}_0 \frac{\partial\beta_d}{\partial \mathbf{\sigma_p}} \\
        \frac {\partial \mathbf{\Ddot{\textbf{ y}}}}{\partial \mathbf{\sigma_p}}& = \frac{1}{2} (P_d \bar n^2 {r_c}^2 -P_d \bar n r_c  \dot{\textbf{y}}_0 ) \frac{\partial\beta_d}{\partial \mathbf{\sigma_p}} \\
         \frac {\partial {\Ddot{ \textbf{z}}}}{\partial \mathbf{\sigma_p}} &= \frac{-1}{2} P_d r_c \bar n \dot{ \textbf{z}}_0 \frac{\partial\beta_d}{\partial \mathbf{\sigma_p}}
\end{align}
\begin{assumption}
The noncooperative debris and the swarm members have similar geometry; therefore $\beta_c = \beta_d$, and $P_d = P_c$ as the state vector approaches zero. This is the cornerstone of the differential drag guidance control law used in this investigation.
\end{assumption}
Some general assumptions are also considered when solving the output regulation as follows.
\begin{assumption}\label{assumption: controllability}
The pairs ($A$,$B$) and ($C$,$A$) are stabilizable and observable, respectively.
\end{assumption}
\begin{assumption}\label{assumption: rank}
{\rm rank}$\left(\begin{bmatrix}
A-\lambda I_n & B\\C&0
\end{bmatrix}\right)=\\n +p,\,\forall\lambda \in \sigma (E)$.
\end{assumption}
In order to solve the optimal output regulation problem, {two optimization} problems need to be addressed. {The static optimization Problem \ref{Problem 1} is solved in order to find the optimal solution $(X^\star,U^\star)$ to the regulator equations \eqref{regeq1}-\eqref{regeq2}. While the dynamic optimization problem described in Problem \ref{Problem 2} is solved to find the optimal feedback control policy.} Both problems are stated as follows:

 \begin{problem}\label{Problem 1}
  \begin{align}
      \min_{(X,U)}{\rm Tr}&(X^{\rm T}\bar{Q}X+U^{\rm T}\bar{R}U), \label{min_Tr}\\
   {\rm subject~ to~~    }  XE&=AX+BU+D,\label{regeq1}\\
     0&=CX+F,\label{regeq2} 
  \end{align}
  where $\bar Q=\left(\bar Q\right)^{\rm T}\succ0$ and $\bar R=\left(\bar R\right)^{\rm T}\succ0$.
   \end{problem}
    \nocite{ABDULLAH20195541}
  Based on Assumption \ref{assumption: rank}, the solvability of the regulator equations defined by \eqref{regeq1}-\eqref{regeq2} is guaranteed and the pair $(X,U)$ exist for any matrices $D$ and $F$; see \cite{huang2004nonlinear}. 
  Additionally, the solution to Problem \ref{Problem 1}, i.e., $(X^\star,U^\star)$ is unique, which will guarantee that the feedforward control policy obtained using $(X^\star,U^\star)$ is also unique and optimal.  
  \begin{problem}\label{Problem 2}
  \begin{align}
      \min_{\bar{u}_i}  \int_{0}^{\infty} &\left(\bar{x}^{\rm T}Q\bar{x}+\bar{u}^{\rm T}{R}\bar{u}\right) ~{\rm d}t, \\\label{eq: error system 1}
      {\rm~subject~ to~~ } \dot{\bar{x}}&=A\bar{x}+B\bar{u},\\\label{eq: error system 2}
      e&=C\bar{x},
  \end{align}
  where $Q=\left(Q\right)^{\rm T} \succeq 0, R =\left(R\right)^{\rm T} \succ0,$ with $\left(A,\sqrt{Q}\right)$ being observable. The eqs \eqref{eq: error system 1}-\eqref{eq: error system 2} form the error system with $\bar{x}:=x-Xv$ and $\bar{u}:=u-Uv$.
  \end{problem}
  
Note that if the deputy's dynamics in \eqref{eq: x-system} are perfectly known, one can develop the optimal controller in the following form.
\begin{align}\label{eq: ui*}
    u^\star(K^\star,L^\star)=-K^\star x+L^\star v,
\end{align} where $K^\star=R^{-1}B^\textrm{T}P^\star$, and $P^\star $ is the unique solution of the following albegraic Ricatti equation (ARE)
\begin{align}
    A^\textrm{T}P^\star+P^\star A+Q-P^\star BR^{-1}B^\textrm{T}P^\star=0.\label{eq: ARE}
\end{align}

The solutions to the regulator equations \eqref{regeq1}-\eqref{regeq2}, i.e., $(X,U)$, form the optimal feedforwad gain matrix such that \begin{align}\label{eq: Li*}
    L^\star&=U+K^\star X.
\end{align}

It is remarkable that equation \eqref{eq: ARE} is nonlinear in $P^\star $. Therefore, different iterative methods have considered to solve the ARE iteratively, including policy iteration (PI) and value iteration (VI). The following lemma shows the convergence of \eqref{eq: ARE} in the sense of the PI method.
\begin{lemma}[\hspace{-0.2pt}\cite{Kleiman1968}]
Let $K_{0} \in \mathbb{R}^{m \times n}$ be a stabilizing feedback gain matrix, the matrix $P_{k}=(P_{k})^{\rm T}\succ0$ be the solution of the following equation  
\begin{align}\label{eq: Policy evaluation}
   \hspace{-1mm} P_{k}(A-BK_{k-1})&+(A-BK_{k-1})^{\rm T} P_{k} +Q\nonumber\\&+K_{k-1}^{\rm T} R K_{k-1}=0,
\end{align} 
 and the control gain matrix $K_{k}$, with $k=1,2,\cdots,$ are defined recursively by
\begin{align}\label{eq: Policy improvement}
    K_{k}=R^{-1} B^\textrm{T} P_{k-1}.
\end{align} 
Then the following properties hold for any $k\in \mathbb{Z}_{+}$. 
\begin{enumerate}
    \item The matrix $A-BK_{k}$ is Hurwitz.
    \item $P^\star \preceq P_{k} \preceq P_{k-1}$.
    \item $\underset{{k\rightarrow \infty}}\lim K_{k} = K^\star ,\; \underset{{k\rightarrow \infty}}\lim P_{k}=P^\star $.
\end{enumerate}
\end{lemma}  

It is notable that an initial stabilizing control policy is required to initiate the learning process of PI. In this paper we consider an iterative reinforcement learning method based on VI to solve $P^\star $, wherein the solvability of the VI is considered under ADP scheme. We consider the use of VI since no initial stabilizing control policy is required to initiate the learning process. This gives VI an advantage over PI since obtaining the prior knowledge of an initial stabilizing control policy is a stringent requirement and may be impossible to obtain, especially when the system dynamics are not available or are not known perfectly. The iterative process of VI to find the optimal control policy is done by repeating the value update step until the value function converges to its optimal value. In the following sections, we show in further details the use of VI to solve our problem.

\section*{Model-Based Value Iteration
}

Throughout this section, the value iteration is used such that the value matrix is iteratively updated until the value matrix converges within a predefined condition.
{To begin with}, $\{B_r\}_{r=0}^{\infty}$ is defined as a collection of nonempty interiors bounded sets, which satisfies 
\begin{align*}
B_r \subset B_{r+1} \in \mathcal{J}_+^n , \; r\in \mathbb{Z}_{+},\; \lim_{r \rightarrow \infty} B_r  = \mathcal{J}^{n}_{+},\end{align*} and $\varepsilon>0$ is a small constant selected as a threshold. In addition, select a deterministic sequence $\{\epsilon_{k}\}_{k=0}^{\infty}$ such that the following conditions are satisfied: \begin{align}\label{eq: ek condition}\epsilon_k>0,\;\;\sum_{k=0}^{\infty}\epsilon_k=\infty,\;\;\lim_{k \rightarrow 0}\epsilon_k = 0.\end{align} 

As mentioned earlier, the VI is different from the policy iteration, described by \eqref{eq: Policy evaluation}-\eqref{eq: Policy improvement} in the sense that an initial stabilizing control policy is not required. Instead, the learning process is initiated with an arbitrary value matrix $P_{0}=(P_{0})^\textrm{T}\succ 0$. In the following, the model-based VI algorithm is given, in which the system matrices $(A,B)$ are used to learn the optimal control policy, based on the results in \cite{Bian2016}.
\begin{algorithm}[H]
\begin{algorithmic}[1]
\caption{Model-based Value Iteration}
\State Select a small constant $\varepsilon>0$, and
{$P_{0}=(P_{0})^\textrm{T}\succ0$}.
\State $k,r \gets 0$.
\Repeat
 \State ${\Tilde{P}_{k+1}\gets P_{k}+\epsilon_k (P_{k}A+A^{\textrm{T}}P_{k}+}$  
 \Statex\hfill$ {  
 Q
 } {-P_{k} B R^{-1} B^{\textrm{T}} P_{k})}$
\If{$\Tilde{P}_{k+1}\notin B_{r}$}
\Statex
{$~~~~~~P_{k+1}\leftarrow P_{0},~ r\leftarrow r+1$}.
\Else 
{ $P_{k+1}\gets\Tilde{P}_{k+1}$} 
\EndIf
\State $k\gets k+1$
\Until $|\Tilde{P}_{k}-P_{k-1}|/\epsilon_{k-1}\prec\varepsilon$
\State $k^\star \gets k$
\State Find the pair $(X,U)$ from \eqref{regeq1}-\eqref{regeq2}.
\State $L_{k\star}\gets U + K_{k^\star}X$
\State Obtain the optimal controller {using} 
\Statex$~~~~~~~~~~~~~~u^\star=-K_{k^\star}x+L_{k^\star}v.$
\label{alg: model-based VI Algorithm}
\end{algorithmic}
\end{algorithm}
\begin{remark}
    It is noteworthy to mention that if the bound of $P^\star$ is known in prior, i.e., $|P^\star|<\gamma$, then one can fix $B_r$ to $B_r=\gamma$. 
\end{remark}

\section*{Data-Driven Value Iteration for Output Regulation Problem}

From the previous section, it is notable that the model-based VI requires the full knowledge of the system matrices $(A,B)$. In practice, obtaining these matrices may not be easy when considering higher order and more complex systems. In this section, we consider a data-driven VI method in which the optimal control policy is obtained without relying on the dynamics or the physics of the system, but the data (state/input information) collected along the trajectories of the underlying dynamical system are used to learn an approximated optimal control policy. 

Considering the $x-$system in \eqref{eq: x-system}, define $\bar{x}_{j}=x - X_{j}v$ for $0\leq j\leq h +1$, where $X_{0}=0_{n \times q}$, $X_{j}\in \mathbb{R}^{n \times q}$ so that $C X_{1}+F=0$. The matrices $X_{j}
$ for $2\leq j\leq h +1$, where $h = (n - p)q$ is the null space dimension of $I_{q}\otimes C$, are selected such that the basis for $\text{ker}(I_{q}\otimes C)$ are formed by all the vectors $\text{vec}(X_{j})$. With the above definitions along with \eqref{eq: exosystem}--\eqref{eq: x-system}, the following differential equation is then obtained.
\begin{align} 
\begin{split} \label{xbar equation for VI}
    \dot{\bar x}_{j}&= A x + B u +(D-X_{j}E)v 
\end{split} \\
\begin{split} \label{xbar equation for PI}
    \dot{\bar x}_{j}&= A_{k} \bar{x}_{j} +B (K_{k} \bar x_{j}+u)\\ 
    &+(D - S (X_{j}))v 
\end{split}
\end{align}
where the Sylvester map $S:\mathbb{R}^{n \times q}\rightarrow \mathbb{R}^{n\times q}$ satisfies $S(X)=XE-AX$, $\forall$ $X\in \mathbb{R}^{n\times q}$, and $A_{k}=A - B K_{k}$.
For any two vectors {$a(t)\in\mathbb{R}^{\boldsymbol{n}},b(t)\in\mathbb{R}^{\boldsymbol{m}}$}, and a sufficiently large $\rho\in\mathbb{Z}_+$, the following matrices are defined.
\begin{align}\label{eq: delta_b}
\delta_b=&\left[\text{vecv}({ b})|_{t_0}^{t_1}~
\text{vecv}({ b})|_{t_1}^{t_2} ~ 
\cdots\right.\nonumber\\
&~~~~~~~~~~~~~~~~~~~~~~~~~~~~~~~~~~~~~~~~~\left.\text{vecv}({ b})|_{t_{\rho -1}}^{t_\rho}\right]^\textrm{T}
,\\
\label{eq: Gamma}
\Gamma_{a,b}=&\left[\int_{t_0}^{t_1}a\otimes b\;\textrm{d}\tau ~ 
\int_{t_1}^{t_2}a\otimes b\;\textrm{d}\tau  
\cdots\right.\nonumber\\ & ~~~~~~~~~~~~~~~~~~~~~~~~~~~~~~~~~~~~~
\left.\int_{t_{\rho-1}}^{t_\rho}a\otimes b\;\textrm{d}\tau\right]^\textrm{T}, \\
\label{eq: I_a}
\mathbb{I}_{a}=&\left[\int_{t_0}^{t_1}\text{vecv}(a)\textrm{d}\tau ~ 
\int_{t_1}^{t_2}\text{vecv}(a)\textrm{d}\tau ~ 
\cdots\right.\nonumber\\& ~~~~~~~~~~~~~~~~~~~~~~~~~~~~~~~~~~~\left. 
\int_{t_{\rho -1}}^{t_\rho}\text{vecv}(a)\textrm{d}\tau\right]^\textrm{T}
.
\end{align}

Consider the Lyapunov candidate $V_{k}(\bar x_{j})=\bar x_{j}^\textrm{T} P_{k} \bar x_{j}$, where $k\in \mathbb{Z}_{+}$. By taking the time derivative of $V_{k}(\bar x_{j})$ along with 
\eqref{xbar equation for PI}, with some mathematical manipulations and rearrangements, one obtains the following.
\begin{align}\begin{split}
    \dot V_k(\bar x_{j})&=\dot{ \bar x}_{j}^\textrm{T}P_{k}\bar x_{j}+\bar x_{j}^\textrm{T}P_{k}\dot{ \bar x}_{j}
\end{split}\nonumber\\
\begin{split} \label{VI Lyapunov}
    &=\bar x_{j}^\textrm{T}(H_{k})\bar x_{j}\\
    &\quad+2u^\textrm{T}RK_{k+1}\bar x_{j}\\
    &\quad+2v^\textrm{T}(D - S(X_{j}))^\textrm{T}P_{k}\bar x_{j}
\end{split}
\end{align}
where $H_{k}=A^\textrm{T}P_{k}+P_{k}A$.\\

By taking the integral of \eqref{VI Lyapunov} over $[t_0, t_s]$, {where $\left\{t_l\right\}_{l=0}^{s}$ with $t_l=t_{l-1}+\Delta t,$ $\Delta t>0$ is a strictly increasing sequence}, the result can be written in the following Kronecker product representation.
\begin{align}\label{VI data-driven}
    &{\Theta_{j}}\begin{bmatrix}
    {\text{vecs}}(H_{k})\\\text{vec}(K_{k+1})\\\text{vec}((D - S(X_{j}))^{\textrm{T}}P_{k})
    \end{bmatrix}=\delta_{\bar x_{j} , \bar x_{j}}\text{vecs} (P_{k})\hspace{-0.2em}
\end{align}
where ${\Theta_{j}} = \begin{bmatrix}
{\mathbb{I}_{\bar{x}_{j}}}
    ,& 2\Gamma_{\bar x_{j},{u}}(I_{n} \otimes R),&2\Gamma_{\bar x_{j},{v}}
    \end{bmatrix} $. If $\Theta_j$ is full column rank, the solution of \eqref{VI data-driven} is obtained in the sense of least square error by using the pseudo-inverse of $\Theta_j$, i.e., $\Theta_j^\dagger=\left(\Theta_j^\textrm{T}\Theta_j\right)^{-1}\Theta_j^\textrm{T}$. The full column rank condition of $\Theta_j$ is satisfied by the following lemma.  
    \begin{lemma}\label{rank lemma}
For all $j\in \mathbb{Z}_{+}$, if there exist a $s'\in \mathbb{Z}_{+}$ such that for all $s>s'$ the following rank condition is satisfied
\begin{equation}
\begin{aligned}
{\rm rank}\left(\begin{bmatrix}
\mathbb{I}_{\bar x_{j}}& 
\Gamma_{\bar x_{j} ,u}&
\Gamma_{\bar x_{j},v}
\end{bmatrix}\right)&=\frac{n (n+1)}{2} +(m+q)n \label{eq: rank}
\end{aligned}
\end{equation}for any increasing sequence $\{t_l\}_{l=0}^{s}$, $t_l=t_{l-1}+\Delta t,$ $\Delta t>0$, then the matrix {$\Theta_{j}$} has full column rank, $\forall\;k\in \mathbb{Z}_{+}.$
\end{lemma}

Lemma \ref{rank lemma} shows that if \eqref{eq: rank} is satisfied, the existence and uniqueness of the solution of \eqref{VI data-driven} is guaranteed, {where the solution can be obtained using the pseudo-inverse of {$\Theta_{j}$}.}
{\begin{remark}
The matrix $\Theta_{j}$ is fixed for all $k\in\mathbb{Z}_+$ and does not require to be updated at each iteration $k$.
\end{remark}}
To this end, the value matrix is updated using stochastic approximation by
\begin{align*}
	{P}_{k+1}\gets P_{k} + \epsilon_k(H_{k} + Q -(K_{k+1})^\textrm{T}RK_{k+1})
\end{align*} where $\epsilon_k$ satisfies \eqref{eq: ek condition}, until the condition $|P_{k}-P_{k-1}|/\epsilon_k\leq{\varepsilon}$ is satisfied, where ${\varepsilon}>0$ is a small threshold. By that, it is guaranteed that the obtained control policy is close enough to the actual optimal one.

 The data-driven VI algorithm can now be introduced. It is presented in Algorithm \label{Algorithm: HI Data-driven}.
\begin{algorithm}[H]
\caption{Data-Driven Value Iteration for Optimal Output Regulation}
\begin{algorithmic}[1]
\State Choose a small threshold constant ${\varepsilon}>0$ and  {$P_{0}=(P_{0})^\textrm{T} \succ 0$}.
\State Compute {the} matrices $X_{0}, X_{1},\cdots, X_{h+1}$.
\State Choose an arbitrary $K_{0}$, not necessarily stabilizing, and employ $u_{0}=-K_{0}x+\eta$, with $\eta$ being an exploration noise over $[t_0,t_s]$.
\State $j\gets 0.$
\Repeat
\State Compute $\mathbb{I}_{\bar{x}_{j}}\text{, }\Gamma_{\bar x_{j}  u} {\text{, and }} \Gamma_{\bar x_{j} v}$ {while} satisfying \eqref{eq: rank}.
\State $j\gets j+1.$
\Until $j=h+2$
\State $k\gets 0$, $j\gets 0$, $r\gets 0$.
\Repeat
\State Solve $H_{k}$ and $K_{k+1}$ from \eqref{VI data-driven}.
\State 
$\Tilde{P}_{k+1}\gets P_{k} + \epsilon_k(H_{k} + Q -(K_{k+1})^\textrm{T}RK_{k+1})$

\If{$\Tilde{P}{_{k+1}}\notin B_r$}
\State {$P_{k+1}\gets P_{0}$, $r\gets r+1$.}\Else
\State  $P_{k+1}\gets \Tilde{P}_{k+1}$
\EndIf \textbf{end if}
\State $k\gets k+1$
\Until $|P_{k}-P_{k-1}|/\epsilon_{k-1}<{\varepsilon}$ 
\State $k\gets k^*, j\gets 1$.
\Repeat
\State From \eqref{VI data-driven}, solve $S({X}_{j})$.
\State $j\gets j+1.$
\Until $j=h +2$

From Problem \ref{Problem 1}, {find} $(X^\star,U^\star)$ using online data.
\State $L_{k^\star}\gets U^\star + K_{k^\star}X^\star$
\State Obtain the suboptimal controller {using} 
\begin{align}\label{eq: suboptimal controller}
    u^\star&=-K_{k^\star}x+L_{k^\star}v.
\end{align}
\end{algorithmic}
\label{Algorithm:HI Data-driven}
\end{algorithm}

If \eqref{eq: rank} is satisfied, it is guaranteed that the sequences $\{{P_{k}}\}_{k=0}^{\infty}$ and $\{{K_{k}}\}_{k=1}^{\infty}$ learned by Algorithm \ref{Algorithm: HI Data-driven} converge respectively to $P^\star$ and $K^\star$. It is worth mentioning that the proposed VI Algorithm \ref{Algorithm: HI Data-driven} is an off-policy learning algorithm. Since the value function in VI is increasing, the increasing sequence of $\{P_k\}_{k=0}^\infty$ will not affect the trajectories of the system during the learning period.

{\begin{remark}
An exploration noise is added to the input of the system \eqref{eq: x-system}--\eqref{eq: e system} during the learning process of Algorithm \ref{Algorithm:HI Data-driven}. Such an input is chosen to satisfy the rank condition \eqref{eq: rank}---which is similar to the condition of persistent excitation. The noise selected can be a random noise or a summation of sinusoidal signals with distinct frequencies, see \cite{sutton2018reinforcement,Bian2016,jiang2012computational} and references therein.
\end{remark}}

\section*{Implementation}

In this section, the simulation of the described scenario using the Basilisk-ROS2 bridge is detailed, and the results are presented.

\subsection*{Basilisk}

Basilisk is an open-source, powerful high-fidelity framework for astrodynamics that combines a message-based architecture and can include multiple programming languages\cite{basi}. One of the advantages of using Basilisk is related to the whole architecture as well as built-in modules, classes, and functions being focused on an Astrodynamics perspective and having a direct correlation to the currently acceptable available literature. For instance, Basilisk's modularity offers the possibility to rapidly modify existing simulation scenarios without fully redesigning the implementation of fundamental concepts, such as spherical or tetrahedral gravitational models, flight controller behavior, and sensor functionality. Moreover, the possibility of fully editing the source code of existing functions and modules creates a significant advantage when using the software framework for novel research development. Furthermore, Basilisk has a fully integrated Unity visualization tool named Vizard, where the user can see animations of the scenario simulated, providing a comprehension beyond traditional plots.

In addition to its noteworthy computation capabilities, the current Basilisk build already accounts for the scenario demonstrated in \textit{Harris et al.} for differential-drag control using a single-facet placed at the geometrical center of the chief and deputy; therefore, to convey a successful implementation, this investigation introduced the above-mentioned J2 model to the Basilisk class \textbf{\textit{simIncludeGravBody}}, and modified the original Modified Rodriguez Parameter Feedback Control module to be able to accept the feedback-forward gains calculated by the data-driven VI algorithm. These modifications, alongside the Basilisk built-in computational tools, will also provide a clear comparison to how the VI controller will perform with respect to a known LQR controller solution. The following is the parameter definitions considered in the simulation:\\
\begin{table}[H]
\centering
\small
\begin{tabular}{|p{3cm}|c|}
\hline
\textbf{Orbital Element} & \textbf{Deputy} \\
\hline
Semi-major Axis (a) & $6678.376~\text{km}$ \\
Inclination (i) & $45~\text{deg}$\\
$\Omega$ & $20.0~\text{deg}$\\
$\omega$ & $30.0~\text{deg}$\\
True Anomaly (f) & $19.75~\text{deg}$\\
\hline
\end{tabular}
\caption{Orbital Elements for Deputy}
\label{tab:orbital_elements_deputy}
\end{table}

\begin{table}[H]
\centering
\small
\begin{tabular}{|p{3cm}|c|}
\hline
\textbf{Orbital Element} & \textbf{Chief} \\
\hline
Semi-major Axis (a) & $6678.136~\text{km}$ \\
Inclination (i) & $45~\text{deg}$\\
$\Omega$ & $20.0~\text{deg}$\\
$\omega$ & $30.0~\text{deg}$\\
True Anomaly (f) & $20~\text{deg}$\\
\hline
\end{tabular}
\caption{Orbital Elements for Chief}
\label{tab:orbital_elements_chief}
\end{table}

\begin{table}[H]
\centering
\small
\begin{tabular}{|p{3cm}|c|}
\hline
\multicolumn{2}{|c|}{\textbf{Spacecraft and Environment Parameters } }\\
\hline
Inertia Tensor (\textbf{I}) & $\text{diag}([10, 9, 8])~\text{kgm}^2$ \\
Mass ($m$) & $6~\text{kg}$\\
$A_i$ & $0.06~\text{m}^2$\\
$C_{D,i}$ & $2.2$\\
$\rho_0$ & $2.2\times10^{-11}~\text{kg/m}^3$\\
$r_c$ & $300~\text{km}$\\
\hline
\end{tabular}
\caption{Scenario Parameters}
\label{tab:scenarioparam}
\end{table}

\subsection*{ROS2 Bridge}

  \begin{figure*}[]
    \centering
    \includegraphics[width = 1\textwidth, height=0.3\textheight
    ]{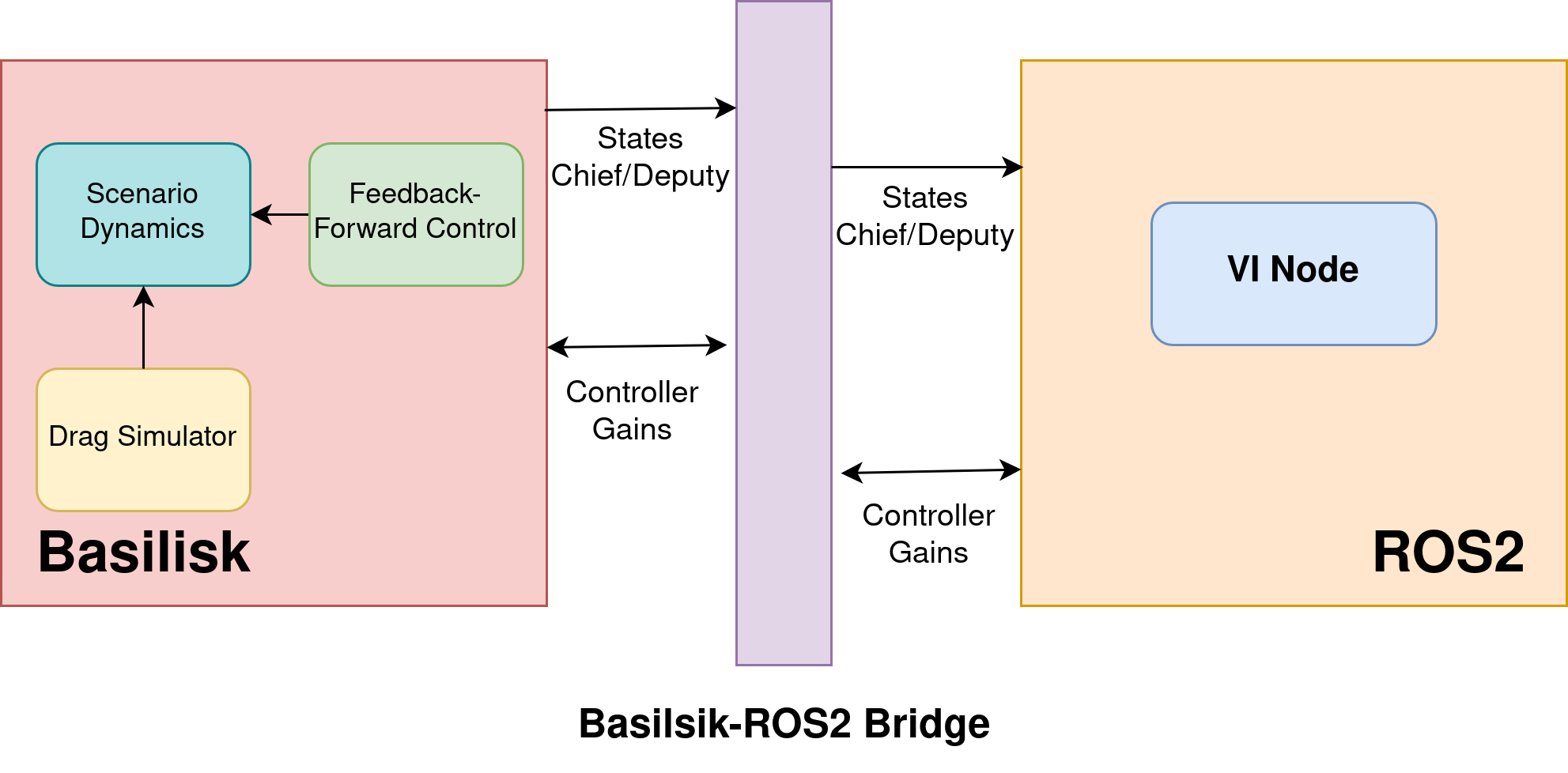}
    \captionsetup{font=small}
    \caption{ROS2-Basilisk Interface}
    \label{fig: Basilisk Interface}
\end{figure*}

\textit{Robotic Operational System} 2 is a multi-field known tool for its capabilities and applications, as it also uses a messaging architecture to communicate between different robotic software segments (nodes). The goal of using ROS2 and Basilisk together relates to the increase in the VI controller application's modularity and the use of Basilisk as a Software-in-the-loop platform for spacecraft applications. The demonstration of the successful Basilisk-ROS2 bridge was introduced by \textit{Matsuka et al}\cite{matsuka2023high}. Although it has some clear limitations in terms of communication delay between the two platforms, the benefits of using a high-fidelity space environment such as Basilisk are significant when considering that generating high-fidelity environments in platforms such as \textit{Gazebo} or \textit{PyBullet} can be extremely time-consuming. It's part of the authors understanding that once the delay between ROS2 and Basilisk can be reduced -- or negligible --, the astrodynamics framework will rise as the best alternative for robotics simulations in space. In this perspective, one can also consider integrating MATLAB instead of Basilisk or in conjunction with the Basilisk-ROS2 bridge, as the \textit{Mathworks} software already has functional integration with ROS2. The former option has clear disadvantages regarding computational speed, Astrodynamics focused modularity, and visualization, given that MATLAB generally has a larger runtime than C/C++ scripts\cite{matcpp} and does not provide a built-in open-source 3D visualization tool. However, the latter option of combining particular well-known and established MATLAB tools, such as the System Identification Toolbox, with the proposed Basilisk-ROS2 integration is viable, but the inherent delay addition remains to be quantified in the literature.   

  \begin{figure}[H]
    \centering
    \includegraphics[width = 1\linewidth, height=0.35\textheight
    ]{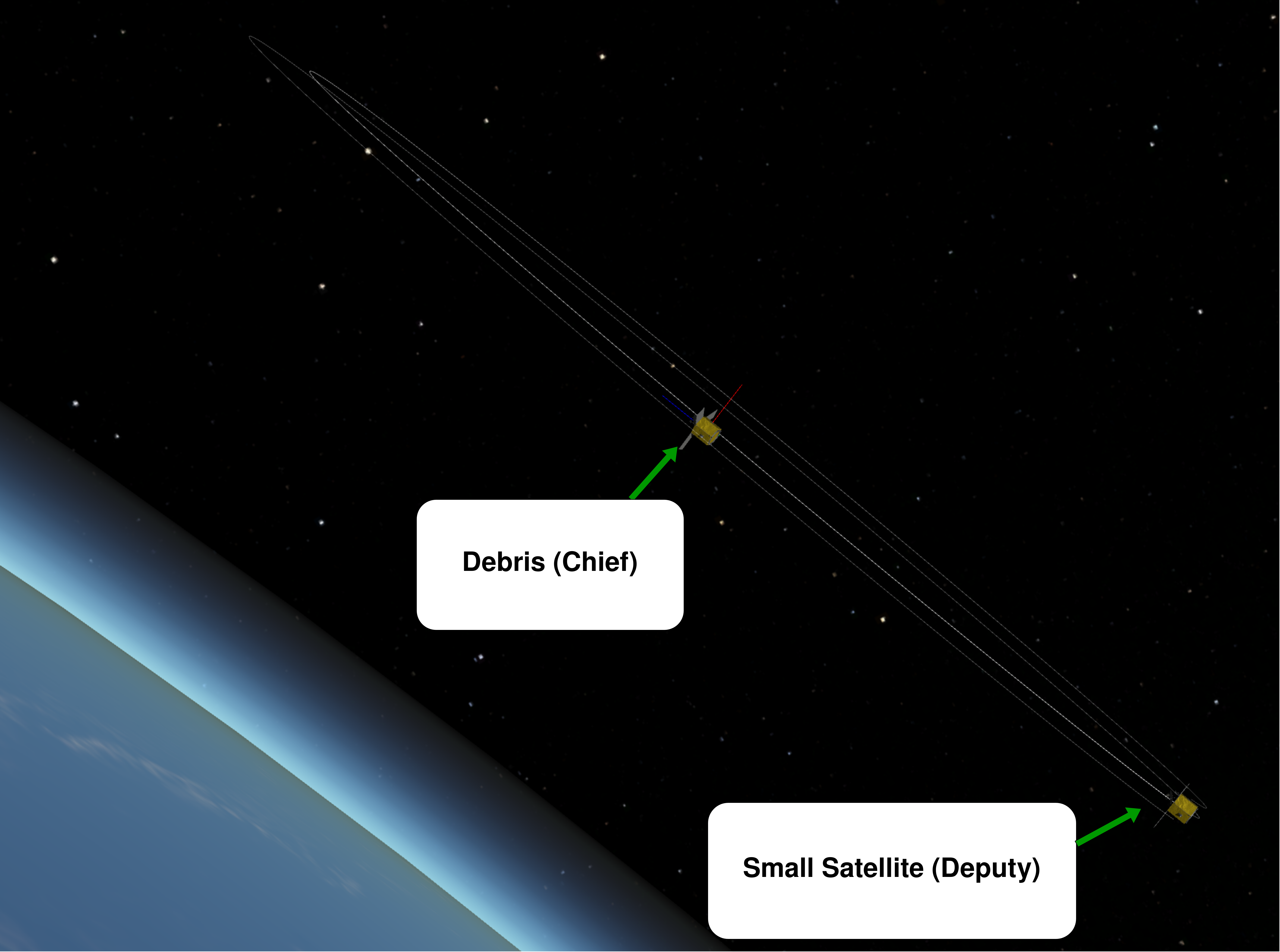}
    \captionsetup{font=small}
    \caption{Scenario Simulated -- Vizard}
    \label{fig: viz}
\end{figure}

As shown in Figure \ref{fig: Basilisk Interface}, one can simulate the entire space environment and system in Basilik and have the current system states transmitted to ROS2, where the data-driven VI controller performs the above-described algorithm and sends the controller gains to Basilisk while and after achieving convergence. Although the process of reaching full convergence could be performed multiple times as new behaviors are detected, e.g., panel actuation failure, this simulation focuses on the ideal case where the system has no unprecedented disturbances apart from J2 oblateness and atmospheric drag. As a result, the VI algorithm only has to reach the full convergence criterion once during the mission. it is noteworthy to rectify that new controller gain messages are beaconed constantly throughout the entire suboptimal to the optimal period only for the data-driven VI. Whereas the LQR gains are considered to be static, meaning that the gain messages are the same throughout the entire mission for the classical model-based approach. A scenario visualization representing the chief and deputy system can be seen in Fig. \ref{fig: viz}. The goal of the simulation is to solve Problem \ref{Problem 1} and Problem \ref{Problem 2} using the classical LQR approach and Algorithm \ref{Algorithm:HI Data-driven} in the context of rendezvous \& capture and formation flying for a deputy/chief system based on differential-drag control.

Therefore, the system dynamics is described in the following form\cite{harris1,Silva2008,schweighart2001development}:
\begin{align}
    &\Ddot{\textbf{x}}-2\bar{n}c\dot{\textbf{y}}-(5c^2-2)\bar{n}^2\textbf{x} + \frac{\beta_d P_d \bar {n} r_c}{2} \dot{\textbf{x}} \nonumber\\
    &=-3\bar{n}^2J_{2}\frac{R_{e}^2}{r_{\rm ref}}\left(\frac{1}{2}-\frac{3\sin^2{(i)} \sin^2{(\bar{n}ct)} }{2}\right.\nonumber\\
    &\quad \left.-\frac{1+3\cos{(2i)}}{8}\right),\label{eq: CW1 J2}
\end{align}


\begin{align}
    &\ddot{\textbf{y}}+2\bar{n}\dot{\textbf{x}} + \beta_d P_d \bar {n} r_c \dot{\textbf{y}} + \frac{\bar{n}^2{r_c}^2(\beta_c P_c - \beta_d P_d)}{2}\nonumber\\
    &= -3\bar{n}^2J_{2}\frac{R_{e}^2}{r_{\rm ref}}\sin^2{(i)}\sin^2{(\bar{n}ct)}\cos{(\bar{n}ct)},\\
    &\ddot{\textbf{z}}+\bar{n}^2\textbf{z} + \frac{\beta_d P_d \bar {n} r_c}{2} \dot{\textbf{z}} \nonumber\\
    &= -3\bar{n}^2J_{2}\frac{R_{e}^2}{r_{\rm ref}}\sin{(i)}\sin{(\bar{n}ct)}\cos{(i)}.\label{eq: CW3 J2}
\end{align}
where $\bar{n}$ is the mean orbital rate, $R_e$ is the radius of the earth, $r_{\rm ref}$ is the position of the reference orbit, $t$ is the time, and $i$ is the angle of incidence.

The system in \eqref{eq: CW1 J2}-\eqref{eq: CW3 J2} can be reformulated and described in the following form: 
\begin{align}
&\Ddot{\textbf{x}} - 2\bar{n}c\dot{\textbf{y}} - (5c^2 - 2)\bar{n}^2\textbf{x} + \frac{\beta_d P_d \bar{n} r_c}{2} \dot{\textbf{x}} \nonumber \\
&= -3\bar{n}^2J_{2}\frac{R_{e}^2}{r_{\rm ref}}\mathcal{Q}_1, \label{eq: xddot} \\
&\ddot{\textbf{y}} + 2\bar{n}\dot{\textbf{x}} + \beta_d P_d \bar{n} r_c \dot{\textbf{y}} + \frac{\bar{n}^2{r_c}^2(\beta_c P_c - \beta_d P_d)}{2} \nonumber \\
&= -3\bar{n}^2J_{2}\frac{R_{e}^2}{r_{\rm ref}}\mathcal{Q}_2, \label{eq: yddot} \\
&\ddot{\textbf{z}} + \bar{n}^2\textbf{z} + \frac{\beta_d P_d \bar{n} r_c}{2} \dot{\textbf{z}}= -3\bar{n}^2J_{2}\frac{R_{e}^2}{r_{\rm ref}}\mathcal{Q}_3. \label{eq: zddot}
\end{align}
where $c\equiv\sqrt{1+\textbf{s}}$ with $\textbf{s}=\frac{3J_2R_e^2}{8{r_{\rm ref}}^2}(1+3\cos{2i})$, $\mathcal{Q}_1$, $\mathcal{Q}_2$ and $\mathcal{Q}_3$ are disturbances generated by the exosystem with $|\mathcal{Q}_1|\leq 1$, $|\mathcal{Q}_2|\leq 1$ and $|\mathcal{Q}_3|\leq 1$. Therefore, the above equations can be transformed in the form of \eqref{eq: exosystem}-\eqref{eq: e system} by assuming sinusoidal signals are generated by the exosystem \eqref{eq: exosystem} in addition to the tracking signals. 
{Based on \eqref{eq: xddot}--\eqref{eq: zddot}, the system matrices can be found as follows:
\begin{align}
A = \begin{bmatrix}
{0}_{3\times 3} & I_{3} \\
\Lambda_1 & \Lambda_2
\end{bmatrix}
\end{align}
where:
\begin{align}
\Lambda_1 = \begin{bmatrix}
(5c^2-2)\bar{n}^2 & 0 & 0 \\
0 & 0 & 0 \\
0 & 0 & \bar{n}^2
\end{bmatrix}
\end{align}
\begin{align}
\Lambda_2 = \begin{bmatrix}
-\frac{1}{2}\beta_d P_d \hat{n} r_c & 2\bar{n}c & 0 \\
-2\bar{n} & -\beta_d P_d \hat{n} r_c & 0 \\
0 & 0 & -\frac{1}{2}\beta_d P_d \hat{n} r_c
\end{bmatrix}
    \end{align}
    \begin{align}
    B &= \begin{bmatrix}
        {0}_{3 \times 3} \\
        {0}_{1 \times 3} \\
        \frac{1}{2}\bar{n}^2 r_c^2 P_d \frac{\partial \beta_d}{\partial \mathbf{\sigma_p}} \\
        {0}_{1 \times 3}
    \end{bmatrix}.
\end{align}

The performed simulation for the data-driven VI is summarized in the following steps:
\begin{enumerate}
\item the simulation time is set to be of $40$ orbital periods $\approx 217247s$
    \item An essentially bounded input is applied to the deputy along with a non-stabilizing control policy.
    \item  State, input, and exosystem information are collected along the trajectories of the system described in \eqref{eq: exosystem}-\eqref{eq: x-system} for the entirety of the mission via Basilisk-ROS2 bridge and parsed into the VI \textit{callback function}.
    \item The differential drag-based optimal control problem is obtained by solving Problem 2, wherein an optimal state feedback gain matrix is obtained.
    \item The output regulation problem is solved by solving Problem 1, wherein the output regulation is then achieved.
    \item The adaptive optimal output regulation is achieved by applying the optimal feedback-feedforward matrix designated in \eqref{eq: Li*} 
\end{enumerate}
The proposed approach shown in Algorithm \ref{Algorithm:HI Data-driven} is used to learn the optimal feedback-feedforward control policy to regulate the relative positions in addition. Instead of using the modeling information of the system, we use the online collected data to learn the optimal control policy, which removes the stringent requirement of knowing the exact physics of the studied system. The data collection and learning are minimally set to be in the time interval from $0$ to $15$ orbital periods. Last but not least, besides achieving the asymptotic tracking of the exosystem signals, we are also able to achieve rejection for class of disturbances generated by the exosystem, with minimizing a predefined cost function. 
{For simulation purposes, we assume the reference signal and the disturbances are generated by the exosystem with the matrix $E$ defined as follows 
\begin{align}
E=\textrm{bdiag}\Bigg(&\begin{bmatrix}
0&0.1\\-0.1&0\end{bmatrix},
\begin{bmatrix}
0&0.2\\-0.2&0\end{bmatrix},\nonumber\\
&\begin{bmatrix}
0&0.3\\-0.3&0\end{bmatrix},
\begin{bmatrix}
0&0.4\\-0.4&0\end{bmatrix}\Bigg)
\end{align}
The rest of the matrices are shown below, where $Dv$ represents the disturbances applied to the system, and $-Fv$ is the tracking signal.
\begin{align}
C&=\begin{bmatrix}
    1&0&0&0&0&0\\
    0&1&0&0&0&0\\
    0&0&1&0&0&0
\end{bmatrix},\\
F&=\begin{bmatrix}
    1&0&0&0&0&0&0&0\\
    0&1&0&0&0&0&0&0\\
    0&0&1&0&0&0&0&0
\end{bmatrix},\\
D&=\begin{bmatrix}
0&0&0&0&0&0&0&0\\
0&0&0&0&0&0&0&0\\
0&0&0&0&0&0&0&0\\
0&0&\Xi&0&0&\Xi&0&0\\
0&0&0&0&\Xi + \xi &0&0&0\\
\Xi&0&0&0&0&0&\Xi&0
\end{bmatrix}
\end{align}
where,\\
$\Xi = -3\bar{n}^2J_{2}\frac{R_{e}^2}{r_{\rm ref}} $, and $\xi =-\frac{\bar{n}^2{r_c}^2(\beta_c P_c - \beta_d P_d)}{2}$ \\ \\
}}
The cost function matrices are considered to be $Q=1.4I_6$ and $R=1\times10^7I_3$. $B_r=10(r+1)$ and $\epsilon_k=\frac{1}{k}$. The CW parameters are chosen same to those used in \cite{schweighart2001development}, where $r_{\textrm{ref}}=6378.136~\text{km}$ and $\Bar{n}=0.00108~1/s$. The altitude of the chief with respect to the earth's center is chosen to be $6678.136 ~\text{km}$ (Low-Earth Orbit).
 Note that matrices $Q$ and $R$ could be modified to mitigate the oscillation experienced in the initial orbital periods, i.e., fast case; However, matrices $Q$ and $R$ were chosen such that the cost function would reach and maintain its minimum for an infinite time horizon, i.e., economic case.

\section*{Results}

\begin{figure}[H]
    \centering
    \includegraphics[width=1\linewidth, height=0.5\textheight, keepaspectratio]{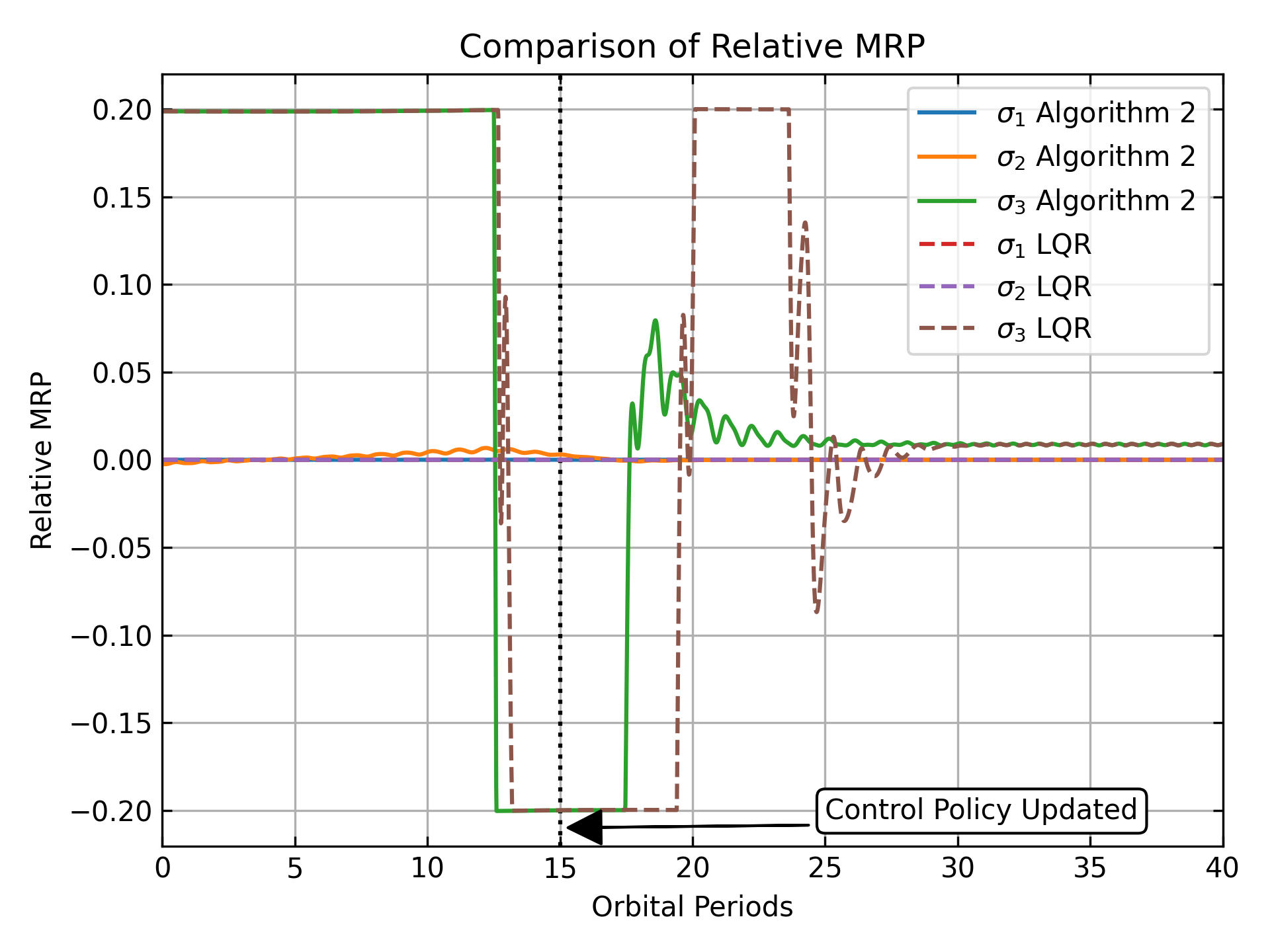}
    \captionsetup{font=small}
    \caption{Relative MRP -- QR Economic Case}
    \label{fig: mrp}
\end{figure}

\begin{figure}[H]
    \centering
    \includegraphics[width=1\linewidth, height=0.5\textheight, keepaspectratio]{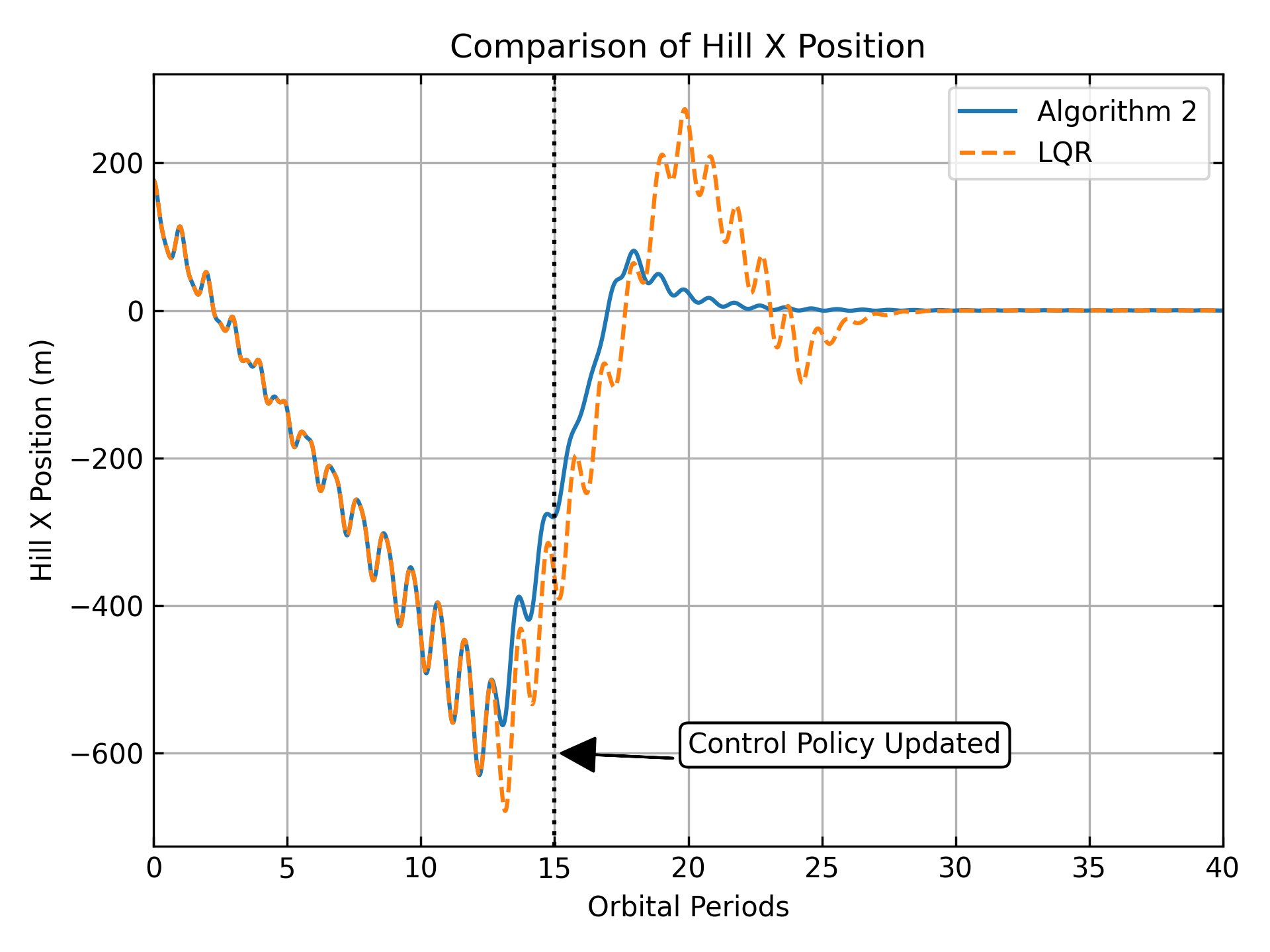}
    \captionsetup{font=small}
    \caption{Hill $\textbf{x}$ Position Comparison -- QR Economic Case}
    \label{fig: hill}
\end{figure}

\begin{figure}[H]
    \centering
    \includegraphics[width=1\linewidth, height=0.5\textheight, keepaspectratio]{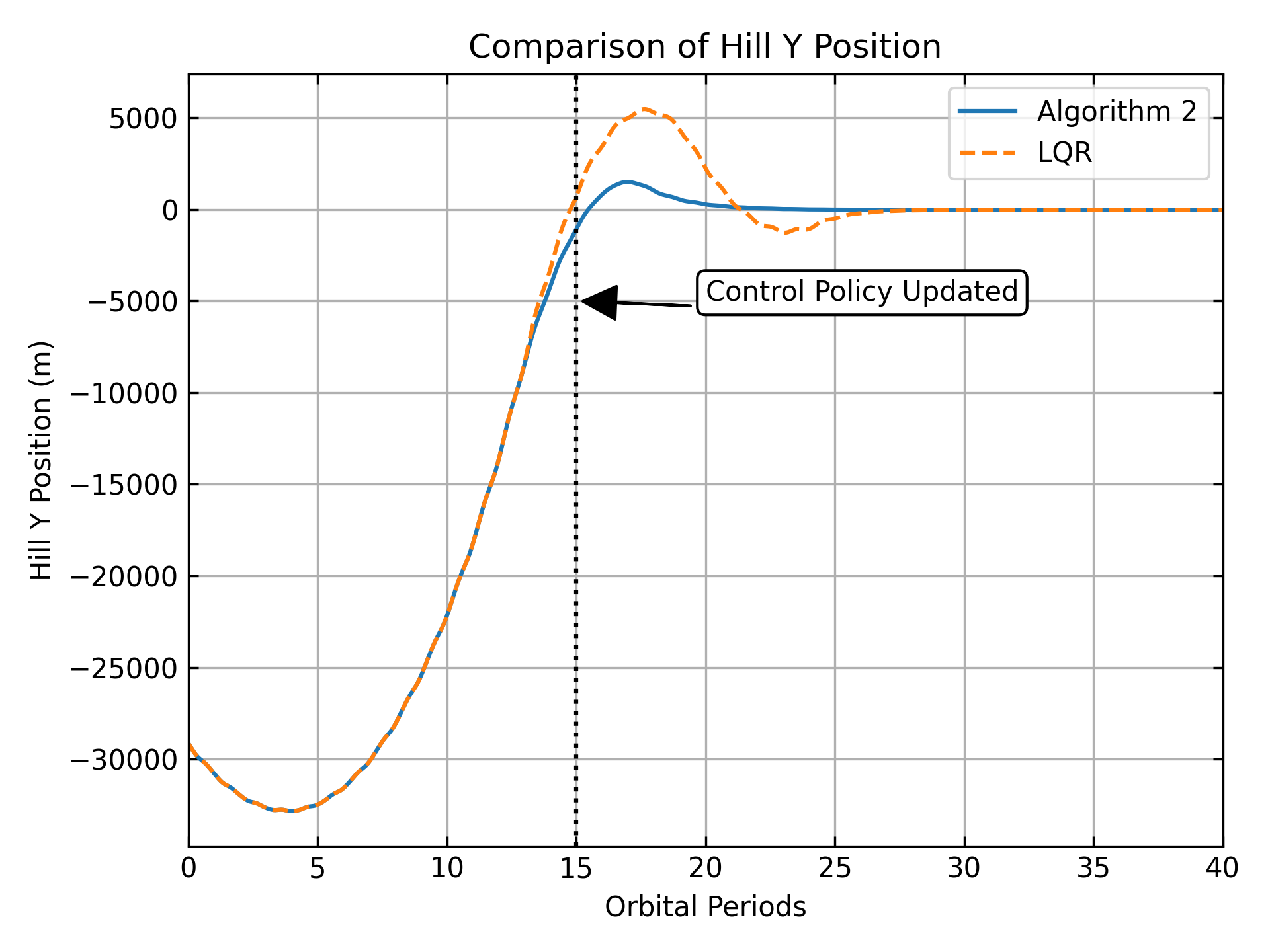}
    \captionsetup{font=small}
    \caption{Hill \textbf{y} Position Comparison -- QR Economic Case}
    \label{fig: hillposY}
\end{figure}

\begin{figure}[H]
    \centering
    \includegraphics[width=1\linewidth, height=0.5\textheight, keepaspectratio]{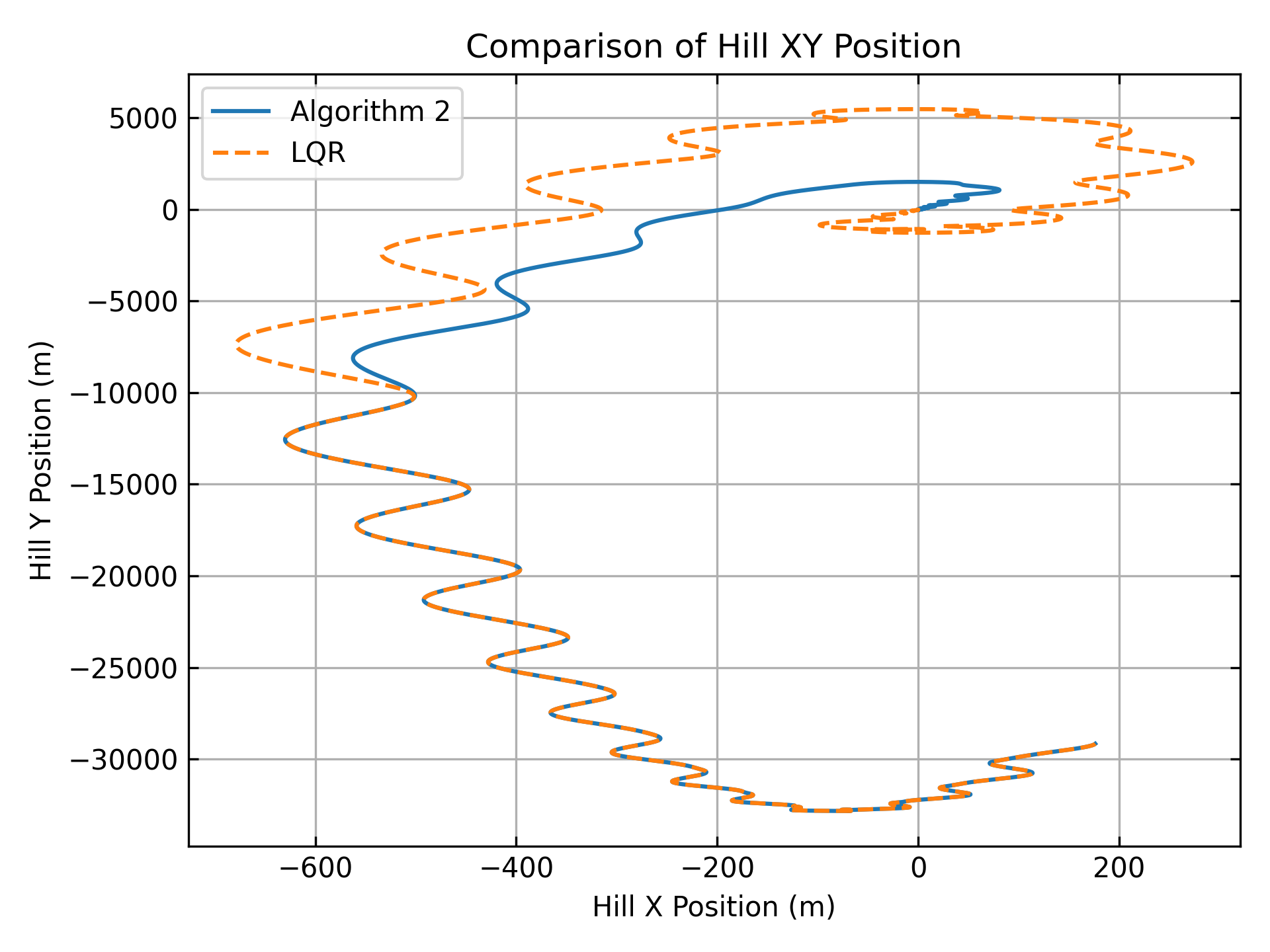}
    \captionsetup{font=small}
    \caption{Hill \textbf{x} vs. \textbf{y} Position Comparison -- QR Economic Case}
    \label{fig: hillposXY}
\end{figure}

\begin{figure}[H]
    \centering
    \includegraphics[width=1\linewidth, height=0.5\textheight, keepaspectratio]{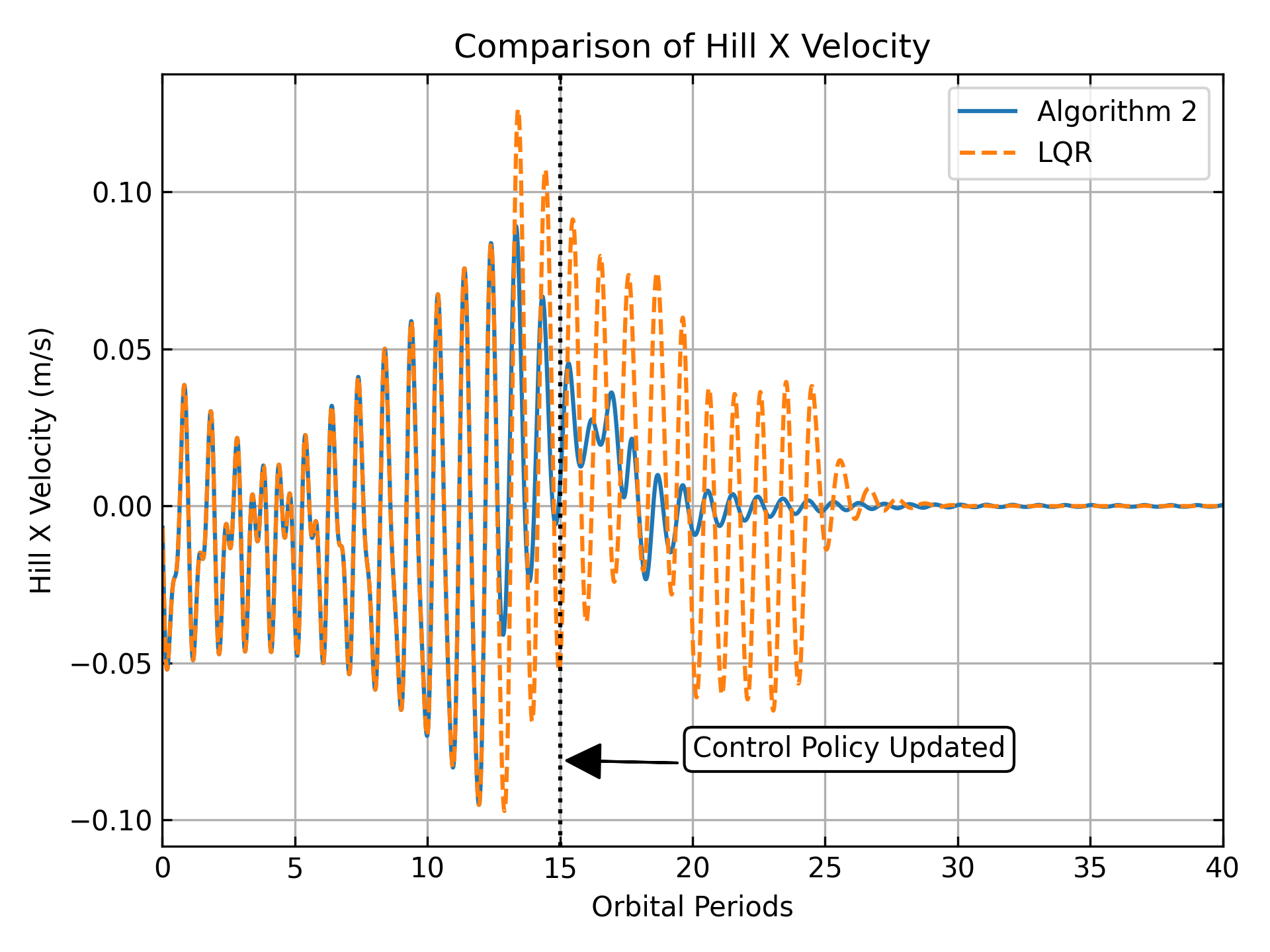}
    \captionsetup{font=small}
    \caption{Hill \textbf{x} Velocity Comparison -- QR Economic Case}
    \label{fig: hillvelx}
\end{figure}

\begin{figure}[H]
    \centering
    \includegraphics[width=1\linewidth, height=0.5\textheight, keepaspectratio]{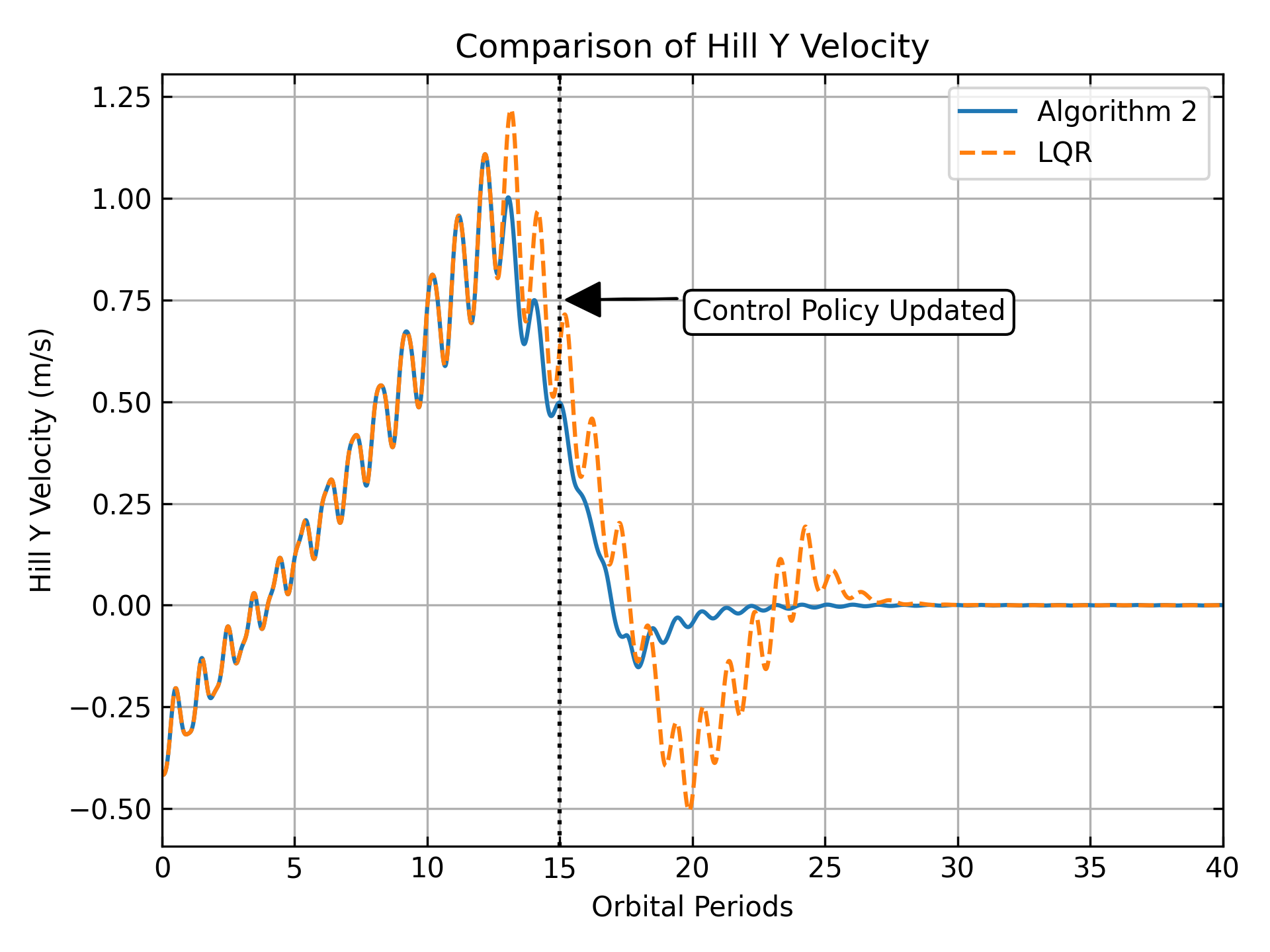}
    \captionsetup{font=small}
    \caption{Hill \textbf{y} Velocity Comparison -- QR Economic Case}
    \label{fig: hillvely}
\end{figure}

For simulation purposes, we assume that the chief (debris) is moving in the space in all $x$, $y$ and that the deputy has an altitude and true anomaly offset, displayed in the Tables \ref{tab:orbital_elements_deputy}-\ref{tab:scenarioparam}. In addition, the velocity in each direction is different. Using Algorithm \ref{Algorithm:HI Data-driven} and static LQR gains and converting their results to the relative attitude equivalent, one can get the feedback-forward gains for the Modified Rodriguez Parameters (MRP) controller. The results obtained are plotted against the orbital period and depicted in Figs. \ref{fig: mrp}-\ref{fig: hillvely}. It's visible that the ROS2-Basilisk adaptive gains approach of the data-driven VI algorithm promotes 
an optimal steady-state and transient performance similar to that obtained when using the LQR gains. 

Fig. \ref{fig: mrp} shows the chief and deputy MRP value difference in each approach, evidencing that the chief and deputy achieve a similar attitude at the end of the maneuver. Furthermore, it is worth mentioning that the oscillations present in Figs. \ref{fig: hill}-\ref{fig: hillvely} are a direct consequence of the choice of matrices $Q$ and $R$ as mentioned in the \textbf{\textit{Implementation}} section.


 It is noticed that the VI approach can achieve convergence in a reasonable amount of time. In addition, it should be highlighted that the computational effort of learning the optimal control policy using VI algorithm is lower than solving the ARE using offline LQR static gains. Moreover, Algorithm 2 removes the requirement of knowing the system matrices $(A,B)$, which is significant for servicing in-orbit small satellites, which may have had their physical structure altered and are non-operational but are still trackable. In addition, the potentially high amount of iterations prior to the VI reaching the full convergence criterion is heavily dependent on the message flow between Basilisk and ROS2; however, when simulating the algorithm offline, the large quantity of iterations is solely due to its sublinear convergence rate and does not necessarily relate to the computational complexity \cite{Bian2016,qasem2023autonomous}. For instance, the PI Algorithm \ref{alg: model-based VI Algorithm} has a quadratic convergence rate but still requires a stabilizing control policy for convergence, increasing the computational complexity \cite{jiang2012computational,QasemTIE2023,QasemCDC2022}.

\section*{Conclusion}

This work considers the control of autonomous formation flying swarm of small satellites in a rendezvous mission with noncooperative trackable debris by developing a direct adaptive optimal control based on differential-drag using adaptive dynamic programming (ADP). More specifically, two problems are considered to guarantee that optimal tracking. First, the output regulation problem is solved to achieve asymptotic tracking and disturbance rejection. Second, we consider solving the output regulation problem by adaptive dynamic programming. Therefore, the states and dynamics information of the system are not needed in order to learn the optimal feedback-feedforward control policy. The ADP approach is implemented on a formation flying and in-orbit servicing problem by considering the Clohessy-Wiltshire equation with J2 perturbations under the effects of atmospheric drag, where the problem is reformulated into an adaptive optimal output regulation problem. The simulation results illustrate the efficacy of the proposed method.

\bibliography{references.bib}{}
\bibliographystyle{unsrt}

\end{document}